\documentstyle[prb,twocolumn,aps]{revtex}
\begin{document}
\draft
\title{Closed-shell interaction in silver and gold chlorides}
\author{
  Klaus Doll$^*$}
\address{
     Max-Planck-Institut f\"ur Physik komplexer Systeme,
          N\"othnitzer Strasse 38, D-01187 Dresden, Germany}
\author{Pekka Pyykk\"o}

\address{
Department of Chemistry, University of Helsinki, POB 55, FIN-00014 Helsinki, 
Finland}
\author{Hermann Stoll}

\address{
Institut f\"ur Theoretische Chemie, Pfaffenwaldring 55, 
Universit\"at Stuttgart, D-70550 Stuttgart,
Germany
}

\maketitle

\begin{abstract}
Hartree-Fock and coupled-cluster calculations have been performed
for cubic AgCl and for AuCl having a cubic or the observed structure with space
group $I4_1/amd$. Cohesive energies and lattice 
constants are in excellent agreement with experiment for AgCl; for AuCl
we find good agreement, and the experimental
structure is correctly predicted to be lower in energy than the cubic one. 
Electron-correlation effects on lattice constants are very large,
of up to 0.8 \AA \ for cubic AuCl.
We especially discuss the strength of the closed-shell interactions, and for
the first time a quantitative analysis of the so-called "aurophilic"
Au(I)-Au(I)  interaction is presented in solids.

\end{abstract}

\pacs{ }

\narrowtext
\section{Introduction}

In recent years, the r\^{o}le of attractive
closed-shell interactions in inorganic chemistry has been 
intensively investigated 
\cite{Krebs} and the results have been summarized in
several review articles \cite{Schmidbaur,Pyykkoe}. 
Both structural, spectroscopic, and energetic evidence exist.
The structures exhibit short secondary M-M distances (M = Cu(I), Ag(I),
Au(I), Tl(I), etc.) and inwards bending of the primary L-M-L' bonds.
The spectroscopic evidence incorporates Raman frequencies of the
M-M stretching mode. The interaction energies can be directly
measured as activation energies by temperature-dependent NMR and 
from optically monitored chemical equilibria.

{\em Ab-initio} theoretical studies exist for intramolecular interactions 
and for dimers. They invariably yield repulsive interactions at
Hartree-Fock level and reproduce the attraction at correlated level. 
Relativistic effects enhance the interaction for gold \cite{PyMe}.
It should be added that the free (Au$^+$)$_2$ dimer remains repulsive   
\cite{LiPyykkoe}. In this sense the ligands are essential.

In this article, we investigate these interactions for the first time in
a solid, using quantum-chemical {\em ab-initio} methods (Hartree-Fock in 
combination with
the coupled-cluster approach). We consider the systems AgCl and AuCl.
While the former has rocksalt structure, the latter one exhibits a rather
complicated tetragonal structure, with space group $I4_{1}/amd$ 
\cite{JanssenAuClpreparation},
which is thermodynamically stable in a region between $\sim$ 90$^\circ$C and 
342$^\circ$C \cite{Janssenphasediagram}. 
This is just an example of a system with relatively short Au-Au distances. 
For the sake of comparison, we performed calculations for both the cubic and the
experimental structure, in the case of AuCl.

\section{The Method}

The calculations consist of two parts. First, Hartree-Fock (HF) calculations
on the solid are performed with
the program package {\sc crystal} \cite{Manual,CRYSTALbuch,PisaniBuch}.
After that, electron correlation for the extended systems is included by means
of a Bethe-Goldstone-like expansion, in terms of correlation
contributions from localized orbital groups ("method of local increments" 
\cite{Stollmethode}).
In a series of studies \cite{Stollmethode,SEMIC,DDFS,Alkali},
it has been shown that this method is capable of yielding results of high
quality. For a detailed account of the method as applied to ionic compounds, 
see Refs.
\onlinecite{DDFS,Alkali}. (In Ref. 
\onlinecite{Alkali}, we also showed how to  determine
van der Waals $C_6$ parameters with this method.)
It should be mentioned that we determined individual increments in finite
model systems (embedded clusters), using an
ionic embedding, with cationic pseudopotentials in the first sphere and point 
charges beyond.
Such an embedding was applied to all systems considered in the present paper,
including AuCl in the experimental structure (which is not fully ionic, 
cf.\ below). 
In order to check the validity of our  
description, we compared the increments taken from 
different clusters (with two or three explicitly described ions) and various
representations of the Madelung field.
The experimental structure turned out to be more sensitive with respect to
the number of point charges than the rocksalt structure, 
so that a larger number is necessary
in the former case. However, the results with
8 $\times$ 8 $\times$ 8 unit cells were nearly identical (up to a few
$\mu$H) to those of 
calculations where only 6 $\times$ 6 $\times$ 6 unit cells were represented
by point charges, which ensures convergence of the results with the larger set.
A more severe problem was the transferability of the increments which was
less good in the observed AuCl structure than in the cubic one.
As correlation scheme, we chose the (size-consistent) coupled-cluster approach
with single and double substitutions (CCSD),
and the CCSD(T) scheme including triples 
by perturbation theory.

For Ag and Au, we used energy-consistent relativistic 19-valence-electron 
pseudopotentials \cite{Andrae}, in combination with the corresponding optimized
$(8s7p6d)$ valence basis sets (uncontracted) and augmented with $3f$ 
and $2g$ functions
(Ag $f/g$-exponents are 2.54, 0.73, 0.20 / 1.58, 0.50; 
Au $f/g$-exponents are 1.41, 0.47, 0.15 /  1.20, 0.40). 
Results for the neutral atoms  and singly charged cations, with these 
pseudopotentials
and basis sets, are shown in 
Table  \ref{AgAuAtom}; the ionization potentials turn out to be in good 
agreement with
experiment. 
For Cl, an all-electron description was chosen; the basis set used for 
the correlation calculations
is a $[6s5p3d2f]$ augmented valence triple zeta basis set \cite{Dunning}
already employed in earlier calculations on alkali halides \cite{Alkali}.
Finally, single-valence-electron pseudopotentials \cite{Fuente}
for Ag and Au were used
as part of the embedding, in our calculations, in order to model the cage 
effect on the explicitly
described cluster atoms (especially the anions).

In the Hartree-Fock calculations, $f$ and $g$ polarization functions as well 
as very diffuse 
functions had to be omitted for Ag and Au. 
The outermost $spd$ exponents were reoptimized 
for the solid. (In the case of AuCl, two different 
basis sets were optimized for the two different crystal structures, 
respectively.) The Cl basis set was taken from Ref. 
\onlinecite{Prencipe}, again reoptimizing the diffuse exponents.
The basis sets for the Hartree-Fock calculations are given in Table 
\ref{CRYSTALbasis}.

\section{Hartree-Fock calculations}
As mentioned, HF self-consistent field (SCF) calculations were performed
with the program package {\sc{crystal}} (Ref. \onlinecite{Manual}).
Results, together with those of the next section, are summarized in Table 
\ref{Sums}.
For AgCl, the SCF cohesive energy, $E_{coh}$, is too small by 30\% with
respect to experiment. This, of course, reflects the well-known SCF error for
bond-breaking. The situation is significantly improved for the lattice energy,
$E_{lat}$, i.e.\ the energy for separating the crystal into Ag$^+$ and Cl$^-$
ions --- however, experiment is still underestimated by 20\%. This has to be 
contrasted
with the situation for the alkali halide crystal NaCl, which has a SCF lattice
constant comparable to AgCl: the percentage of the SCF errors for both 
$E_{coh}$ and $E_{lat}$ are smaller by roughly 10-15\% \cite{Alkali,Prencipe}.
This should not be taken as an indication of increased covalent 
contributions in AgCl ---
the Mulliken population analysis shows a still nearly perfect
ionic behaviour, with charges of $\pm$ 0.9 $e$ for Ag and Cl.
However, it is certainly an indication that the correlation contribution 
of the Ag$^+$
core is much more important than for the alkali halides. A similar 
conclusion can
be drawn from considering the lattice constant: for AgCl, SCF yields a 
deviation
from experiment of 0.5 \AA \  (9\%) --- in contrast  to 'normal' quantum-chemical
SCF results for covalent systems, it is an overestimation. This points to the
importance of bond-shortening correlation effects like dynamic polarization 
and dispersion
in AgCl. Of course, these effects are also acting in the alkali 
halides, but their magnitude
is significantly smaller (3.5\% in NaCl). 
Note that this is in line with the trend of the 'in-crystal' polarizabilities 
of the cations Na$^+$ and Ag$^+$ to be discussed below.
The SCF bulk modulus of AgCl ($\sim$ 23 GPa)
grossly underestimates experiment, by more than 50\%.

Calculations for AuCl in the cubic rocksalt structure
yield a SCF lattice constant and a SCF lattice energy both very similar to 
the corresponding
values for AgCl.
Also, according to our population analysis, the charges in AuCl
are $\pm$0.9 $e$, i.e.\ essentially the same as in AgCl.
Thus, it may be concluded that the closed-shell repulsive potentials
are quite similar for the (relativistic) Ag$^+$ and Au$^+$ ions.
The cohesive energy of AuCl, on the other hand, is smaller by
more than 1 eV than in AgCl.
This can be explained by the relativistic stabilization
of the 6$s$ electron in the free Au atom \cite{HayWadtKahnBobrowicz}.
Turning now to our results for the experimental tetragonal ($I4_1/amd$) structure of AuCl,
we find it to be energetically nearly degenerate to the cubic structure, at 
the SCF level, with the latter
very slightly lower (by $\sim$1 mH).
However, one can not conclude that at the Hartree-Fock
limit (i.e.\ with a complete basis set) the experimental structure would 
still be higher in energy. 
In the rocksalt structure, the $d$-occupancy on the chlorine atom is
only 0.01 $e$ and omitting the $d$-function decreases the cohesive
energy by only 2 mH,
whereas the $d$-population in the experimental structure is 0.05 $e$ and the 
energy decreases by 17 mH without this function. 
This results from the fact that a static $d$-polarization is symmetry-forbidden
in the rocksalt structure \cite{Prencipe}.
Thus, one might expect that additional polarization functions 
(e.g. $f$ functions on Au)
should further lower the energy of the experimental structure
compared to the cubic one, and then already at the HF limit the 
experimentally
observed structure would become more stable.
As already indicated by the stronger static polarization effects, we find a
tendency towards more covalent bonding in the experimental structure: this
shows up in the Mulliken population analysis, where the charge transfer Au
$\rightarrow$ Cl is found to be not more than 0.5 $e$. In line with this 
change in bonding,
the SCF overestimation of the lattice constant (5.8\% for the experimental 
structure)
is smaller than for the cubic one. (Note, at this point, that
we performed only calculations for the isotropically expanded 
experimental structure, i.e. the ratio $\frac{a}{c}$ was kept fixed as well
as the parameter $z=0.19$ which characterizes the $z$-component of the chlorine
atoms with respect to the gold atoms.)

For AgCl, Hartree-Fock and density functional calculations
have been performed \cite{Apra,Geipel}. The Hartree-Fock results
are similar to ours (i.e. about 70 \% of the cohesive energy
is recovered and the lattice constant is overestimated by 0.4-0.5 \AA). In
Ref. \onlinecite{Geipel}, a detailed analysis of relativistic 
contributions was made,
and  reductions of the bond length, of about 1 \%, and of the
cohesive energy, of about 15 mH, were found when effects of relativity were
included. Correlation-energy density-functional corrections improve the HF 
results 
significantly; however, 
the lattice constant is still overestimated \cite{Apra}.
An {\em ab-initio} treatment of electron correlation in silver and gold halides
has not been attempted so far.

\section{Correlation calculations}

The calculations were done with the program package {\sc molpro}
\cite{KnowlesWerner,MOLPROpapers}. The correlated orbitals are
$3s$ and $3p$ for Cl; for Ag and Au we correlated the highest occupied
$d$-shell (i.e.\ 4$d$ for Ag, 5$d$ for Au). We found in CCSD calculations 
for AgCl that additionally correlating the inner-core $4s$ and $4p$ orbitals
only slightly changes the results (e.g., the Ag-Cl increment for next neighbors
changes from 0.008077 H to 0.008576 H).
The incremental expansion is made in terms of localized orbital groups
which can be attributed either to the metal cations, M$^+$, or are mainly
centered on the chlorine anions. However, this does not imply any {\em a 
priori}
assumption with regard to the ionicity of the system.

In Tables \ref{AgCl} and \ref{AuClNaCl}, we give results for 
AgCl and AuCl (cubic structure) at two different lattice constants;
corresponding results for the tetragonal structure of AuCl are
contained in Table \ref{AuClexp}.
Let us first consider the
 intra-ionic 'one-body' contributions which are obtained
when correlating only one embedded ion. As can be seen from the
Tables, these
energies are not too different from those of the free ions and therefore
make only a minor contribution to the lattice energy.
Much more important are the van der Waals like inter-ionic
two-body contributions; these
are determined by correlating a pair of ions and subtracting off the 
corresponding intra-ionic one-body terms.
We begin the discussion with results for the rocksalt structure. 
The dominant two-body contributions clearly are
the metal-chlorine terms. This is to be expected since these ions are nearest
neighbors and the polarizability is not only high for the Cl$^-$ ions 
but also for the metal ions due to their
large number of electrons and large ionic radii. The chlorine-chlorine
contributions are smaller by a factor of 4 ...\ 5 because of the larger 
distance between nearest-neighbor
Cl ions in the lattice, but in turn are still significantly more important 
(by a factor of 3 ...\ 4)
than the metal-metal ones. It may be interesting to compare the AgCl 
correlation-energy increments to those of NaCl
which has about the same lattice constant. There, the Cl-Cl increments 
are the dominant ones, with the Na-Cl
terms smaller by $\sim$20 \% and the Na-Na ones completely negligible --- 
of course, this just reflects the
relation between the metal-ion dipole polarizabilities $\alpha$: that of 
Ag$^+$ being only a factor 2 smaller than for Cl$^-$,
$\alpha$(Na$^+$), on the other hand, being smaller than $\alpha$(Cl$^-$) by 
more than an order of magnitude.
It has been argued \cite{Bucher} that the difference in the strength of van 
der Waals interactions is responsible
for the unusual features in the cohesive properties of AgCl as compared to 
the alkali halides.
In fact, according to our present CCSD(T) calculations (and those of Ref.\ 
\onlinecite{Alkali}), the ratio
of inter-ionic correlation contributions in AgCl and NaCl, at the equilibrium 
lattice constants, is $\sim$4.4,
which is not very different from the ratio of $\sim$6 estimated by Bucher 
\cite{Bucher} on the basis of a semiempirical
fit to solid-state data.
Comparing now inter-ionic correlation in AgCl and AuCl, one notes an increase 
AgCl $\rightarrow$ AuCl by about
a factor 2.  Of course, this has to do with the increase 
of the metal $\alpha$ but, as shown by the
increase of the Cl-Cl increments,
is also influenced by the reduction of the lattice constant in AuCl. In spite 
of the increased importance of 
inter-ionic van der Waals interaction in AuCl, the Au-Au increments are still 
rather small in absolute value:
for a given Au-Au pair of nearest neighbors, the effect is only $\sim$0.04 eV.
Let us now consider the experimental tetragonal AuCl structure.
Here, distances between the ions are shorter, with a lower coordination 
number, however (smaller weight factors).
This means that, for example, the Au-Cl two-body increment for next 
neighbors is much larger than in the rocksalt structure, but has to be multiplied
with a weight factor of two only (six in the cubic structure). The sum
is still clearly dominated by the Au-Cl contributions which in total
are roughly the same (within $\sim$15\%) as in the cubic structure, the same holds for the
Cl-Cl increments. In contrast to that, the  
correlation contribution between Au ions increases by a factor of $\sim$3. 
The largest van der Waals attraction between
an individual Au-Au pair (or, to be more precise, between the two $d^{10}$ 
shells)
is now $\sim$0.2 eV, i.e.\ relatively strong (comparable to the values 
given in Figure 36 of Ref. \onlinecite{Pyykkoe} at this distance of 3.2 \AA).
However, this attraction alone would certainly not be strong enough to 
overcompensate 
the electrostatic repulsion of the positively charged ions (even if a 
reduced charge of
0.5 $e$, according to the Mulliken population analysis, is assumed). We have no 
indication,
furthermore, for an additional stabilizing covalent interaction between the 
Au ions:
the SCF overlap population is -0.02. On the other hand, the $6s$-like 
valence population
on Au could give rise to non-negligible valence (and core-valence) correlation
contributions --- only, we cannot separate these contributions from those 
of the
Cl ions, since they cannot be attributed to different localized orbitals.

In Table \ref{vdWTabelle}, we display 'in-crystal' ionization potentials 
(IP), dipole
polarizabilities ($\alpha$) and $C_6$ coefficients
for the cubic structures. 
The IP and $\alpha$ were calculated for an embedded single ion 
(treated with high-quality basis set and
surrounded by pseudopotentials/point charges). For evaluating $C_6$, we 
multiplied individual
2-body increments, for a given species, involving ions with the smallest 
internuclear distance
in the lattice, by the sixth power of their distance. Of course, this 
implies the assumption of
a pure van der Waals interaction between these ions, which is certainly 
only approximately satisfied.
As already mentioned, the in-crystal polarizability of Ag is larger by a factor 
$\sim$10 than that of Na, i.e.
essentially comparable to Rb \cite{Alkali}, 
and the $\alpha$ value of Au is even higher 
by a factor $\sim$1.5.
The $C_6$ coefficients for Cl-Cl in AgCl, and even more so in AuCl, 
are larger than for the alkali halides
\cite{Alkali}. While this trend can be qualitatively explained for AgCl 
using the London formula, cf.\
 Table \ref{vdWTabelle}, the rationalization for AuCl is more difficult. 
Apparently, the spill-over of
charge on neighboring crystal ions is larger in AuCl, and the interaction is 
less van der Waals-like.
The $C_6$ coefficients for metal-chlorine interaction are next highest in 
magnitude.
That of AgCl is in good agreement with RbCl, as expected from the similarity 
of the polarizabilities,
and the change from AgCl to AuCl roughly scales with the metal 
polarizabilities again.
Turning now to the interactions between metal ions:
our CCSD $C_6$ value for Ag-Ag (71 a.u.) is in
the range of 62 to 375 a.u. given in Ref. \onlinecite{WilsonMadden}.
The Au-Au $C_6$ coefficient (149 a.u.) is certainly very high compared to other
metal-metal coefficients because of the high polarizability, 
but still significantly lower than both Au-Cl and Cl-Cl.
This coefficient is also much lower when comparing to results from
literature for $C_6$ coefficients characterizing the van der Waals 
interaction between two molecules, for example for the dimer (H$_3$P-Au-Cl)$_2$
a long-distance $C_6$ limiting 
value of 1830 a.u. has been calculated \cite{PyMe}.

In order to check whether or not intra-ionic and 2-body inter-ionic 
correlation-energy
increments give a reliable account of correlation effects in the systems 
considered,
we also calculated several three-body increments. 
They were determined, as non-additivity corrections, in (embedded) 
clusters with three
explicitly correlated ions, cf.\ footnotes to Tables
\ref{AgCl}, \ref{AuClNaCl}, and \ref{AuClexp}.
As in earlier work they do not make an important contribution to the
total energy.
This does not seem to support the argument of Ref. 
\onlinecite{Bucher} that three-body van der Waals
interactions are important in AgCl. However, it is not clear whether the 
violation of
the Cauchy relation of the elastic constants in AgCl would show up 
explicitly by means
of 3-body increments in our formalism, or rather by an (angular) dependence of
the 2-body ones on changes of lattice geometry.

Our final results are obtained by adding the (weighted) sum over all 
correlation-energy
increments to the SCF energies. By doing this for a range of lattice 
constants, we can
monitor the influence of correlation on lattice constants and bulk 
moduli (see Table
\ref{Sums}). While HF cohesive energies
underestimate the experimental values by 30 \% (AgCl) or 50 \% (AuCl),
the agreement is excellent at the correlated level 
for AgCl with deviations of less than 2\%. For AuCl, we find a less good 
agreement with an overestimation of 34 mH (17 \%) for the cohesive energy
or 30 mH (7\%) for the lattice energy.
Lattice constants are
strongly reduced due to the van der Waals interaction, by up to 0.8 \AA \ 
or 14 \% (CCSD(T))
in the hypothetical cubic structure of AuCl. In the rocksalt structure, AuCl would
have a shorter lattice constant (by 0.3 \AA) than AgCl.
The shortest Au-Cl distances found in the tetragonal structure of AuCl 
are 2.36 \AA, the shortest
Au-Au and Cl-Cl distances are 3.22 \AA \mbox{ } and 3.21 \AA \mbox{ }, respectively 
(our calculated 
CCSD(T) values are 3.9 \% shorter). 
The reduction of the distances
compared to the cubic structure (CCSD(T) values: Au-Au = Cl-Cl = 3.69 \AA  , 
Au-Cl =  2.61 \AA )
may be explained with the reduced ionic radii in systems with 
smaller coordination number \cite{Shannon}.
The larger deviation of our results for the experimental tetragonal AuCl 
structure as compared to the cubic AgCl one is probably due
to the less good transferability of the increments in the former structure
(see footnotes to 
Tables \ref{AgCl}, \ref{AuClNaCl}, and \ref{AuClexp}). When we extract
the largest Au-Cl increment (which makes the most important correlation contribution
to the lattice energy) from a cluster with three (instead of two)
explicitly described ions, the cohesive energy is reduced by 12.5 mH and
the deviation of the lattice constant from experiment shrinks to
3.5 \%, at the CCSD(T) level. 
Note also that
our calculations refer to zero temperature where AuCl is experimentally
unstable.
While the cubic structure for AuCl was slightly lower in energy (by 1 mH) 
than the
experimental one, at the SCF level, the situation is reversed by correlation,
and the experimental structure is calculated to be significantly more stable 
now (by 39 mH).
About half of the stabilization is due to Au$^+$-Cl$^-$ pairs, the next
important contribution ($\sim$40\%) coming from the van der Waals interaction of the
Au$^+$ cores. Both effects involve correlation of the Au $5d$ shell.

\section{Conclusion}

We have extended the method of local increments to $4d$ and $5d$ transition-metal compounds.
The results for cohesive energy and geometry are in excellent agreement
with experiment for AgCl and in good agreement for AuCl. 
For AuCl, we find the experimental structure with space group $I4_1/amd$
to be lower in energy by $\sim$ 
39 mH than the hypothetical rocksalt structure. While
the latter structure corresponds to nearly purely ionic bonding, the 
experimental structure features significant
covalent bonding. The largest correlation contributions to the lattice 
energy come from pairs of nearest-neighbor
metal/halide ions (up to $\gtrsim$1 
eV for a Au-Cl pair in the experimental structure of AuCl).
However, while the metal-metal pairs give only very small van der Waals-like 
contributions in the cubic
structure of AgCl and AuCl ($\leq$0.04 eV for a single M-M interaction), 
the correlation effect of
Au-Au pairs reaches that of the Cl-Cl ones ($\sim$0.2 eV per pair) in the 
tetragonal structure.
The stabilization of the tetragonal structure with respect to the cubic one
is essentially a correlation effect involving excitations from the Au $5d$ shells.
The influence of correlation on the geometry turns out to be very important:
the lattice constant is reduced by 9 \% for AgCl and 10 \%
AuCl (experimental structure), for AuCl in the 
rocksalt structure the reduction would be even larger (14 \%).

\acknowledgments
KD and HS are grateful to Prof. P. Fulde (Dresden) for continous
support. KD would like to thank Daresbury Laboratory where this work was
completed.
\\ \\
$^*$ Present  address:
Daresbury Laboratory, Daresbury, Warrington, WA4 4AD, Great Britain

\onecolumn

\begin{table}
\begin{center}
\caption{\label{AgAuAtom}The first ionization potential (IP) for Ag and 
Au is given in Hartrees units, evaluated at Hartree-Fock and correlated levels,
in comparison with experimental values.}
\vspace{5mm}
\begin{tabular}{ccccccc}
Atom & RHF & \multicolumn{1}{c}{CCSD} & \multicolumn{1}{c}{CCSD(T)} & 
expt.\cite{Moore}\\
\hline
Ag & 0.233121 & 0.269033 &  0.272983 & 0.27842 \\
Au & 0.282265 & 0.326956 &  0.331273 & 0.33904 \\
\end{tabular}
\end{center}
\end{table}

\begin{table}
\begin{center}
\caption{\label{CRYSTALbasis}
Exponents and contraction coefficients (in parentheses)
of the basis sets for the {\sc crystal} 
calculation. In the case of AuCl, we optimized two different basis sets for
the cubic and the experimental structure.} 
\vspace{5mm}
\begin{tabular}{ccccccccc}
 &  & Exponent (contraction) & & & \multicolumn{2}{c}{Exponent (contraction)}\\
\multicolumn{3}{c}{AgCl} & \multicolumn{4}{c}{AuCl} \\ \\
  & & & & & cubic lattice & experimental lattice\\
\hline
Ag& $1s$, 2$sp$, 3$spd$ & pseudopotential \cite{Andrae} 
&Au & 1$s$, 2$sp$, 3$spd$, 4$spdf$ & \multicolumn{2}{c}{pseudopotential 
\cite{Andrae}}
 \\ 
  & $4s$ & 9.0884420 (-1.9648132) &  & 5$s$ & \multicolumn{2}{c}{20.1152990 (-0.1597614)}  \\
 &    &  7.5407310 (2.7332194) & & & \multicolumn{2}{c}{12.1934770  (0.7893559)} \\
 &    &  2.7940050 (0.1991148) & & & \multicolumn{2}{c}{6.0396260 (-1.5714057)} \\
 & $5s$ & 1.4801580 & & $6s$ & \multicolumn{2}{c}{1.3737210} \\
 & $6s$ & 0.65& & $7s$ & 0.65001 & 0.630 \\
 & $7s$ & 0.16 & & $8s$ &  0.171 & 0.102\\
 & $4p$ & 4.4512400 (-6.0833780) & & $5p$ & \multicolumn{2}{c}{8.6096650 (2.0982231)}\\ 
 &    & 3.6752630 (6.4168543) & & & \multicolumn{2}{c}{7.3353260 (-3.0458670)}\\
 & $5p$ & 1.2912880  (0.7539735) & & 6$p$ & \multicolumn{2}{c}{1.9129700 (0.3791452)} \\
      &    & 0.6525780 (0.2730597) & & & \multicolumn{2}{c}{1.0576950 (0.6456428)} \\
      & 6$p$ &     0.38                & & 7$p$ & 0.452 & 0.442 \\ 
      & 4$d$ & 7.9947300  (-0.0163876) &  & 5$d$ &  \multicolumn{2}{c}{4.1439490 (-0.4058458)} 
\\
      &    &       2.7847730   (0.2814107)  & &  & \multicolumn{2}{c}{3.5682570   (0.4275070)} 
\\
      &    &  1.2097440   (0.4863264)  & & & \multicolumn{2}{c}{1.3443240 (0.4755405)} \\
      &    &  0.5053930   (0.3867258) & & & \multicolumn{2}{c}{0.5552890 (0.5610972)} \\
      & 5$d$ & 0.198851  & & 6$d$ & 0.192 & 0.188 \\
   Cl & $1s$, 2$sp$, 3$sp$ & as in Ref. \onlinecite{Prencipe}&
   Cl &      1$s$, 2$sp$, 3$sp$ & \multicolumn{2}{c}{as in Ref. 
   \onlinecite{Prencipe}} \\
      & 4$sp$ & 0.308 & & 4$sp$ & 0.314 & 0.312 \\
      & 5$sp$ & 0.113 & & 5$sp$ & 0.115 & 0.108\\
      & 3$d$ & - & & 3$d$ & 0.46 & 0.40 \\
\end{tabular}
\end{center}
\end{table}

\begin{table}
\begin{center}
\caption{\label{Sums}Final results for AgCl and AuCl. Energies are given in Hartree. } 
\vspace{5mm}
\begin{tabular}{ccccccccc}
 & & HF & CCSD & CCSD(T) & exp. \\
\hline
 AgCl& & & & \\
 & E$_{coh}$ & 0.1421 & 0.1885 & 0.1997 & 0.2031 \cite{CRC} \\
 & E$_{lat}$ & 0.2801 & 0.3315 & 0.3414 &  0.3487 \\
 & $a$ & 6.00 \AA &  5.57 \AA  &      5.52 \AA  & 5.51 \AA \cite{Apra} \\
 & B & 23 GPa & 47 GPa & 50 GPa &   53.5 $\pm$ 0.5 GPa  \cite{Apra} \\
\\
AuCl (cubic structure) \\
 & E$_{coh}$ & 0.0994 & 0.1730 & 0.1936 & - \\
 & E$_{lat}$ & 0.2865 & 0.3738 & 0.3956 & - \\
 & $a$ & 6.04 \AA & 5.35 \AA     &  5.22 \AA & - \\
\\
 AuCl (exp. structure) \\
 & E$_{coh}$ & 0.0986 & 0.2067 & 0.2328 & 0.1989 \cite{CRC} \\
 & E$_{lat}$ & 0.2858 & 0.4076 & 0.4348 & 0.4051 \\
 & $a/c$       & exp. +5.8 \% &
exp. -3.1 \%  & exp. -3.9 \%  &  6.734 \AA / 8.674 \AA 
\cite{JanssenAuClpreparation}  \\
\end{tabular} 
\end{center}
\end{table}

\onecolumn
\begin{table}
\begin{center}
\caption{\label{AgCl}Local correlation-energy increments 
(in Hartree)
for cubic AgCl.
The increments include weight factors appropriate to one formula unit of the crystal.
Position vectors (i,j,k) are given in units of $a/2$, where $a$ is the lattice constant.} 
\vspace{5mm}
\begin{tabular}{ccccccccc}
& & \multicolumn{4}{c}{lattice constant} & \\
 & & \multicolumn{2}{c}{5.55 \AA} & \multicolumn{2}{c}{6.00 \AA} \\
 & weight factor & CCSD & CCSD(T) & CCSD & CCSD(T) \\
\hline
embedded Cl$^-$ & 1 & -0.219821 & -0.228511 & -0.220706 & -0.229654 \\
free Cl$^-$     & -1 & +0.222073 & +0.231560 & +0.222073 & +0.231560 \\
embedded Ag$^+$ & 1 & -0.344594 & -0.357267 & -0.344572 & -0.357224 \\
free Ag$^+$ & -1 & +0.344551 & +0.357182 & +0.344551 & +0.357182 \\
\rm{Cl(0,0,0)-Cl(0,1,1)} & 6 & -0.010217 & -0.012170 & -0.006733 & -0.008127\\
\rm{Cl(0,0,0)-Cl(2,0,0)} & 3 & -0.000455 & -0.000546 & -0.000299 & -0.000362\\
\rm{Cl(0,0,0)-Cl(2,1,1)} & 12 & -0.000482 & -0.000580 & -0.000324 & -0.000393\\
\rm{Ag(0,0,0)-Cl(1,0,0)} & 6 & -0.048464 & -0.056076 & -0.032905 & -0.038437 \\
\rm{Ag(0,0,0)-Cl(1,1,1)} & 8 & -0.001714 & -0.002040 & -0.001033 & -0.001234 \\
\rm{Ag(0,0,0)-Cl(2,1,0)} & 24 & -0.000899 & -0.001069 & -0.000566 & -0.000675\\
\rm{Ag(0,0,0)-Ag(0,1,1)} & 6 & -0.002565 & -0.003000 & -0.001475 & -0.001724 \\
\hline
 {one-body-contribution to} &  & +0.002209 & +0.002964 & +0.001346 & +0.001864 
\\ 
 lattice energy \\
 two-body-contribution to & & -0.064769 & -0.075481 & -0.043335 & -0.050952 \\
 lattice energy \\
 total correlation contribution & 
& -0.062587 & -0.072517 & -0.041989 & -0.049088
\\
 to lattice energy$^a$
\end{tabular}
\end{center}
$^a$ We calculated two examples for three-body increments at 5.55 \AA: 
The increment
Cl(1,0,0)-Cl(0,1,0)-Cl(0,0,1) is +0.000041 H, at the CCSD level (CCSD(T): +0.000043);
the CCSD increment for Ag(0,1,0)-Cl(0,0,0)-Cl(0,1,1) is +0.000103 H (CCSD(T): +0.000071 H).
When taken from the latter cluster, CCSD results for one- and two-body increments are:
Ag$^+$: -0.344138 H, Cl$^-$: -0.219889 H,
 Ag-Cl: -0.048006 H, Cl-Cl: -0.011010 H;
comparison with the corresponding values of the Table shows that the transferability of the
increments is good.
\end{table}

\onecolumn
\begin{table}
\begin{center}
\caption{\label{AuClNaCl}Local correlation-energy increments
(in Hartree)
for cubic AuCl. 
The increments include weight factors appropriate to one formula unit of the crystal.
Position vectors (i,j,k) are given in units of $a/2$, where $a$ is the lattice constant.} 
\vspace{5mm}
\begin{tabular}{ccccccccc}
& & \multicolumn{4}{c}{lattice constant} & \\
 & & \multicolumn{2}{c}{5.35 \AA} & \multicolumn{2}{c}{6.00 \AA} \\
 & weight factor & CCSD & CCSD(T) & CCSD & CCSD(T) \\
\hline
embedded Cl$^-$ & 1 & -0.216956 & -0.225227 & -0.218147 & -0.226611 \\
free Cl$^-$     & -1 & +0.222073 & +0.231560 & +0.222073 & +0.231560 \\
embedded Au$^+$ & 1 & -0.316662 & -0.329529 & -0.316643 & -0.329463 \\
free Au$^+$ & -1 & +0.316628 & +0.329420 & +0.316628 & +0.329420 \\
\rm{Cl(0,0,0)-Cl(0,1,1)} & 6 & -0.016820 & -0.020159 & -0.007916 & -0.009572\\
\rm{Cl(0,0,0)-Cl(2,0,0)} & 3 & -0.001359 & -0.001604 & -0.000460 & -0.000549\\
\rm{Cl(0,0,0)-Cl(2,1,1)} & 12 & -0.000697 & -0.000849 & -0.000302 & -0.000365\\
\rm{Au(0,0,0)-Cl(1,0,0)} & 6 & -0.094949 & -0.110352 & -0.053367 & -0.062973 \\
\rm{Au(0,0,0)-Cl(1,1,1)} & 8 & -0.003991 & -0.004798 & -0.001517 & -0.001818 \\
\rm{Au(0,0,0)-Cl(2,1,0)} & 24 & -0.001851 & -0.002220 & -0.000767 & -0.000916\\
\rm{Au(0,0,0)-Au(0,1,1)} & 6 & -0.006699 & -0.007879 & -0.002897 & -0.003404 \\
\hline
 {one-body-contribution to} &  & +0.005083 & +0.006224 & +0.003911 & +0.004906 
\\ 
 lattice energy \\
 two-body-contribution to & & -0.126366 & -0.147861 & -0.067224 & -0.079597 \\
 lattice energy \\
 total correlation contribution & 
& -0.121283 & -0.141637 & -0.063313 & -0.074691 \\
 to lattice energy$^a$
\end{tabular}
\end{center}
$^a$ One three-body increment was calculated at 5.55 \AA: 
the CCSD contribution for Au(0,1,0)-Cl(0,0,0)-Cl(0,1,1) is +0.000233 H (CCSD(T): +0.000172 H).
CCSD results for one- and two-body increments, when taken from this cluster, are:
Au$^+$: -0.315803 (-0.316652) H, Cl$^-$: -0.217860 (-0.217040) H, Au-Cl: -0.075516 (-0.079530) 
H, 
Cl-Cl: -0.015054 (-0.013038 H). Comparison with the numbers in parentheses, 
which correspond
to results from clusters with one or two explicitly described ions, 
shows again that the transferability is good.
\end{table}

\onecolumn
\begin{table}
\begin{center}
\caption{\label{AuClexp}Local correlation-energy increments
(in Hartree)
for AuCl (experimental structure). $\vec a_{exp}$ represents the experimental 
lattice constants $a$=6.734 \AA \mbox{ } and $c$=8.674 \AA.
The increments include weight factors appropriate to one formula unit of the 
crystal.} 
\vspace{5mm}
\begin{tabular}{ccccccccc}
& & \multicolumn{4}{c}{lattice constant} & \\
 & & \multicolumn{2}{c}{0.95$*\vec a_{exp}$} & \multicolumn{2}{c}{1.00$*
\vec a_{exp}$} \\
 & weight factor & CCSD & CCSD(T) & CCSD & CCSD(T) \\
\hline
embedded Cl$^-$ & 1 & -0.219282 & -0.228022 & -0.221509 & -0.230664 \\
free Cl$^-$     & -1 & +0.222073 & +0.231560 & +0.222073 & +0.231560 \\
embedded Au$^+$ & 1 & -0.318164 & -0.331708 & -0.318760 & -0.332342 \\
free Au$^+$ & -1 & +0.316628 & +0.329420 & +0.316628 & +0.329420 \\
\rm{Cl(0,0.25,0.19)-Cl(0,0.25,0.56)} & 1/2 & -0.002949 & -0.003472 & -0.002257 & -0.002678 \\
\rm{Cl(0,0.25,0.19)-Cl(0.5,0.25,0.31)} & 1 & -0.004004 & -0.004782 & -0.003016  & -0.003627 \\
\rm{Cl(0,0.25,0.19)-Cl(0.5,0.25,-0.06)} & 2 & -0.006536 & -0.008034 & -0.004654 & -0.005802 \\
\rm{Cl(0,0.25,0.19)-Cl(0,0.75,-0.19)} & 1 & -0.004097 & -0.004923 & -0.002514 & -0.003070 \\
\rm{Cl(0,0.25,0.19)-Cl(0.5,0.75,0.06)} & 2 & -0.003042 & -0.003814 & -0.002000 & -0.002556 \\
\rm{Cl(0,0.25,0.19)-Cl(0,0.25,-0.44)} & 1/2 & -0.000731 & -0.000950 & -0.000451 & -0.000596 \\
\rm{Au(0,0,0)-Cl(0,0.25,0.19)} & 2 & -0.080882 & -0.092820 & -0.070456 & -0.081532 \\
\rm{Au(0,0,0)-Cl(0.5,0.25,-0.06)} & 4 & -0.022088 & -0.027312 & -0.016036 & -0.020052 \\
\rm{Au(0,0,0)-Cl(0,0.25,-0.44)} & 2 & -0.010300 & -0.012722 & -0.007608 & -0.009576 \\
\rm{Au(0,0,0)-Cl(0.5,0.25,0.31)} & 4 & -0.002424 & -0.002952 & -0.001676 & -0.002044 \\
\rm{Au(0,0,0)-Cl(0,0.25,0.56)} & 2 & -0.000482 & -0.000578 & -0.000342 & -0.000410 \\
\rm{Au(0,0,0)-Au(0.25,0.25,-0.25)} & 2 & -0.014226 & -0.017044 & -0.009960 & -0.011908\\
\rm{Au(0,0,0)-Au(0,0.5,0)} & 1 & -0.004624 & -0.005527 & -0.003183 & -0.003799 \\
\rm{Cl(0,0.25,0.19)-Au(0,0.5,0)-Cl(0,0.75,-0.19)}$^a$ & 1 & +0.001329 & +0.001231 &
+0.001133 & +0.001075 \\
\hline
 {one-body-contribution to} &  & +0.001255 & +0.001250 & -0.001568 & -0.002026\\ 
 lattice energy \\
 two-body-contribution to & & -0.156385 & -0.184930 & -0.124153 & -0.147650 \\
 lattice energy \\
 total correlation contribution & 
& -0.153801 & -0.182449 & -0.124588 & -0.148601 \\
 to lattice energy
\end{tabular}
\end{center}
$^a$ Following CCSD results for one- and two-body increments
have been extracted from the above three-atom cluster at 
1.00$*\vec a_{exp}$: 
Cl: -0.220213 H, Au: -0.316332 H, Cl(0,0.25,0.19)-Cl(0,0.75,-0.19): -0.003499 H,
Au(0,0,0)-Cl (0,0.25,0.19): -0.061924 H. Comparison with the corresponding values of
the Table shows that the transferability is less good than in the cubic structure, especially 
for the
Au-Cl increment between next neighbors which makes the most important contribution
to the lattice energy.

\end{table}

\begin{table}
\begin{center}
\caption{\label{vdWTabelle} C$_6$ coefficients, determined from CCSD
two-body increments. In-crystal
polarizabilities ($\alpha$) and ionization potentials (IP) were calculated as
in Ref. 14. The last column is a comparison with an 
estimate from London's formula. All quantities except for lattice constants
are in atomic units.}
\vspace{5mm}
\begin{tabular}{ccccccccc} 
 System & lattice & \multicolumn{1}{c}{$\Delta E$}  &
 $-C_{6}=\Delta E$ $\times$ $r^6$ & IP$_{cat}$ & IP$_{an}$ &
$\alpha_{cat}$ & $\alpha_{an}$ & $ -\frac{2}{3}\frac{r^6}{\alpha_{1}
\alpha_{2}}\frac{IP_1 + IP_2}{IP_1 IP_2}$ $\times$ $\Delta E$\\
& constant $a$ in \AA &  & & & & & & \\
\hline
Ag-Cl (AgCl) & 5.55 & -0.008077 & 170 & 0.46 & 0.44 & 9.3 & 21.5 & 2.5 \\
Ag-Ag (AgCl) &      & -0.000428 &  71 & & & & & 2.4 \\
Cl-Cl (AgCl) &      & -0.001703 & 283 & & & & & 1.9 \\
Au-Cl (AuCl) & 5.35 & -0.015825 & 264 & 0.42 & 0.36 & 13.4 & 20.5 & 3.3 \\
Au-Au (AuCl) &      & -0.001117 & 149 & & & & & 2.6 \\
Cl-Cl (AuCl) &      & -0.002803 & 374 & & & & & 3.3 \\
\end{tabular}
\end{center}
\end{table}


\begin{references}
\bibitem{Krebs}
{\it Unkonventionelle Wechselwirkungen in der Chemie metallischer
Elemente: Bericht zum Schwerpunktprogramm "Neue Ph\"anomene in der
Chemie metallischer Elemente mit abgeschlossenen inneren 
Elektronenzust\"anden"}
Deutsche Forschungsgemeinschaft, Editor: B. Krebs (VCH, Weinheim, 1991).
\bibitem{Schmidbaur}H. Schmidbaur, Gold Bull. {\bf 23}, 11 (1990).
\bibitem{Pyykkoe} P. Pyykk\"o, Chem. Rev. {\bf 97}, 597 (1997).
\bibitem{PyMe} P. Pyykk\"o, N. Runeberg, and F. Mendizabal, Chem. Eur. J.
{\bf 3}, 1451 (1997); P. Pyykk\"o and F. Mendizabal, Chem. Eur. J.
{\bf 3}, 1458 (1997); P. Pyykk\"o and F. Mendizabal, Inorg. Chem. (submitted).
\bibitem{LiPyykkoe} J. Li and P. Pyykk\"o, Chem. Phys. Lett.
{\bf 197}, 586 (1992).
\bibitem{JanssenAuClpreparation}E. M. W. Janssen, J. C. W. Folmer, and
G. A. Wiegers, J. Less-Common Metals {\bf 38}, 71 (1974);
J. Str\"ahle and K.-P. L\"orcher, Z. Naturforsch. {\bf 29b}, 266 (1974).
\bibitem{Janssenphasediagram} E. M. W. Janssen, F. Pohlmann, and G. A. Wiegers,
J. Less-Common Metals {\bf 45}, 261 (1976).
\bibitem{Manual} R. Dovesi, V. R. Saunders, and C. Roetti, M. Caus\`a, 
N. M. Harrison, R. Orlando, and E. Apr\`a, {\sc crystal 95} User's
Manual, Theoretical Chemistry Group, University of Torino (1996).
\bibitem{CRYSTALbuch} C. Pisani, R. Dovesi, and C. Roetti,
{\em Hartree-Fock Ab Initio Treatment of Crystalline Systems}, edited by 
G. Berthier et al, Lecture Notes in Chemistry Vol. 48 (Springer, Berlin,
1988).
\bibitem{PisaniBuch} {\em Quantum-Mechanical Ab-initio Calculation of the
Properties of Crystalline Materials}, Editor: C. Pisani (Springer, Berlin,
1996).
\bibitem{Stollmethode} H. Stoll, Phys. Rev. B {\bf 46}, 6700 (1992);
H. Stoll, Chem. Phys. Lett. {\bf 191}, 548 (1992).
\bibitem{SEMIC}
B. Paulus, P. Fulde, H. Stoll, Phys. Rev. B {\bf 51}, 10572 (1995);
B. Paulus, P. Fulde, H. Stoll, Phys. Rev. B {\bf 54}, 2556 (1996);
S. Kalvoda, B. Paulus, P. Fulde, H. Stoll, Phys. Rev. B {\bf 55}, 4027 (1997);
B. Paulus, F.-J. Shi, H. Stoll, J. Phys.: Condens. Matter {\bf 9}, 2745
(1997); M. Albrecht, B. Paulus, and H. Stoll, Phys. Rev. B {\bf 56},
7339 (1997).
\bibitem{DDFS}K. Doll, M. Dolg, P. Fulde, H. Stoll, Phys. Rev. B {\bf 52},
4842 (1995); K. Doll, M. Dolg, H. Stoll, Phys. Rev. B {\bf 54}, 13529 (1996);
K. Doll, M. Dolg, P. Fulde, H. Stoll, Phys. Rev. B {\bf 55}, 10282 (1997);
K. Doll, M. Dolg, P. Fulde, H. Stoll, Chemical Papers {\bf 51}, 357 (1997).
\bibitem{Alkali} K. Doll and H. Stoll, Phys. Rev. B {\bf 56}, 10121 (1997);
Phys. Rev. B {\bf 57}, 4327 (1998).
\bibitem{Andrae}D. Andrae, U. H\"au{\ss}ermann, M. Dolg, H. Stoll, and
H. Preu{\ss}, Theor. Chim. Acta {\bf 77}, 123 (1990).
\bibitem{Dunning} D. E. Woon and T. H. Dunning, Jr., J. Chem. Phys. {\bf 98},
1358 (1993).
\bibitem{Fuente} P. Fuentealba, H. Stoll, L. v. Szentp\'aly, P. Schwerdtfeger, 
and H. Preu{\ss},
J. Phys. B {\bf 16}, L323 (1983).
\bibitem{HayWadtKahnBobrowicz}P. J. Hay, W. R. Wadt, L. R. Kahn, and 
F. W. Bobrowicz, J. Chem. Phys. {\bf 69}, 984 (1978).
\bibitem{Prencipe} M. Prencipe, A. Zupan, R. Dovesi, E. Apr\`a, and V. R.
Saunders, Phys. Rev. B {\bf 51}, 3391 (1995).
\bibitem{Apra} E. Apr\`a, E. Stefanovich, R. Dovesi, and C. Roetti,
Chem. Phys. Lett. {\bf 186}, 329 (1991); Chem. Phys. Lett. {\bf 197}, 338
(1992) (Erratum).
\bibitem{Geipel} N. J. M. Geipel and B. A. He{\ss}, Chem. Phys. Lett. 
{\bf 273}, 62 (1997).
\bibitem{KnowlesWerner}{\sc molpro} is a package of {\em ab-initio} programs 
written
by H.-J. Werner and P.J. Knowles, with
contributions from J. Alml\"of, R. D. Amos, M. J. O. Deegan, F. Eckert,
S. T. Elbert, C. Hampel, W. Meyer, A. Nicklass, K. Peterson,
R. M. Pitzer, A. J. Stone, P. R. Taylor, M. E. Mura, P. Pulay, M. Schuetz,
H. Stoll, T. Thorsteinsson, and D. L. Cooper.
\bibitem{MOLPROpapers}
H.-J. Werner and P.J. Knowles, J. Chem. Phys. {\bf 82}, 5053 (1985);
P.J. Knowles and H.-J. Werner, Chem. Phys. Lett. {\bf 115}, 259 (1985);
H.-J. Werner and P.J. Knowles, J. Chem. Phys. {\bf 89}, 5803 (1988);
P.J. Knowles and H.-J. Werner, Chem. Phys. Lett. {\bf 145}, 514 (1988);
C. Hampel, K. Peterson, and H.-J. Werner, Chem. Phys. Lett. {\bf 190},
1 (1992);
P.J. Knowles, C. Hampel, and H.-J. Werner, J. Chem. Phys. {\bf 99}, 5219
(1993); M.J.O. Deegan and P.J. Knowles, Chem. Phys. Lett. {\bf 227}, 
321 (1994).
\bibitem{Bucher} M. Bucher, Phys. Rev. B {\bf 30}, 947 (1984); Phys. Rev. B
{\bf 35}, 6432 (1987).
\bibitem{WilsonMadden}M. Wilson, P. A. Madden, and B. J. Costa-Cabral,
J. Phys. Chem. {\bf 100}, 1227 (1996).
\bibitem{Shannon} R. D. Shannon, Acta Cryst. A {\bf 32}, 751 (1976).
\bibitem{Moore} C. E. Moore, Atomic Energy Levels, NSRDS-NBS 35 / Vol. I-III,
Nat. Bur. Standards (Washington, DC, 1949, 1952, 1958).
\bibitem{CRC} CRC Handbook of Chemistry and Physics, 75$^{th}$ edition,
Editor: David R. Lide (CRC Press, Boca Raton, 1994/1995).






















\end{references}
\end{document}